# Self-organized Archimedean Spiral Pattern: Regular Bundling of Fullerene through Solvent Evaporation

Yong-Jun Chen[1,2,3,a], Kosuke Suzuki[2], Hitoshi Mahara[2], Kenichi Yoshikawa[3,a], Tomohiko Yamaguchi[2,a]

[1]Department of Physics, Shaoxing University, Shaoxing, Zhejiang Province 312000, China

[2]Nanosystem Research Institute, National Institute of Advanced Industrial Science and Technology (AIST), 1-1-1 Higashi, Tsukuba, Ibaraki 305-8565, Japan

[3]Faculty of Life and Medical Sciences, Doshisha University, Kyotanabe, Kyoto, 610-394, Japan

We report the spontaneous generation of an Archimedean spiral pattern of fullerene via the evaporation of solvent. The self-organized spiral pattern exhibited equi-spacing on the order of μm between neighboring stripes. The characteristics of the spirals, such as the spacing between stripes, the number of stripes and the band width of stripes, could be controlled by tuning the thickness of the liquid bridge and the concentration of solution. The mechanism of pattern formation is interpreted in terms of a specific traveling wave on the liquid-solid interface accompanied by a stick-slip process of the contact line.

a) To whom correspondence should be addressed. Electronic address: chenyongjun@usx.edu.cn (Y. J. C), keyoshik@mail.doshisha.ac.jp (K. Y.), tomo.yamaguchi@aist.go.jp (T. Y.).

Recently, regular structures that are generated through the deposition of various building blocks, such as colloidal particles [1, 2], nanowires [3], nanoparticles [4], quantum dots [5] and macromolecules [6], have been attracting considerable interest as a unique method for surface treatment and the fabrication of micro-devices [7-9]. Various methods of surface patterning have been developed for the "bottom-up" fabrication of functional structures [10-17]. Precipitation via the evaporation of solutions containing the aforementioned building blocks near the contact line at a meniscus produces regular periodic-stripe patterns. Dispersed solutes have been shown to be transferred to the vicinity of the contact line by capillary flow and then self-assembled into stripe patterns by a stick-slip cycle [6]. In addition, a gradient stripe pattern has been found in the evaporation process of a confined solution [3, 6]. However, most of these previous studies, including both experimental and theoretical works, considered the stripes as concentric rings. Promising theoretical studies, while few [18], have focused on the morphology of rings or lines without considering other morphologies, such as spirals. Additionally, only patterns with a gradient with regard to the stripe separation have been reported in most experimental works. The presence of a spatial gradient or inhomogeneity in the concentric pattern is not favorable for the fabrication of an ordered structure for practical applications [8, 9]. The further development of the methods for producing controllable and programmable structures (for example, with a specific number of stripes in the structure, an adjustable space between neighboring rings and a suitable width of the stripes) is indispensable. In this letter, we demonstrate the generation of an Archimedean spiral pattern from deposition following the evaporation of a confined solution, which exhibits equi-spacing with a desired periodicity on the order of a micrometer. This spiral morphology of stripes represents a mode of self-organized pattern formation.

We used Fullerene C60 (99.5%, Aldrich) and toluene (Wako Chemicals). Fullerene C60 was dissolved in toluene in an ultrasonic bath (temperature: 9-14℃). Glass beads (diameter: 1.73~1.77 mm) were confined between two parallel glass plates (Fig. 1a) and situated on the lower plate. Fullerene C60 solution was placed between the two parallel glass plates to fill the confined space, as shown in Fig. 1. The gap between the top of the glass beads and the surface of the upper glass plate was adjusted from 0 mm to 0.260 mm. The confined solution was then placed in a vacuum chamber and the internal pressure was gradually decreased (pressure>0.65kPa). Air that had adsorbed on the surfaces of the glass beads and glass plates was carefully removed. As toluene evaporated, a precipitation pattern of fullerene C60 crystals was generated on both the upper and lower glass plates. The pattern was observed under a laser scanning microscope (KEYENCE VK 9710) and analyzed with image-analysis software. The experiments were performed at room temperature (20 ℃).

Figure 1(b) shows a typical self-organized spiral pattern on the surface of the upper glass plate (the superposition line is shown in the inset of Fig. 1(c)). The spiral stripes have a nearly uniform width, and a similar height, as revealed by a three-dimensional profile of the pattern (not shown). As shown in Fig. 1(c), a plot of the radius ($R$) of the spiral stripe shows linear dependence on the angle ($\theta$) as

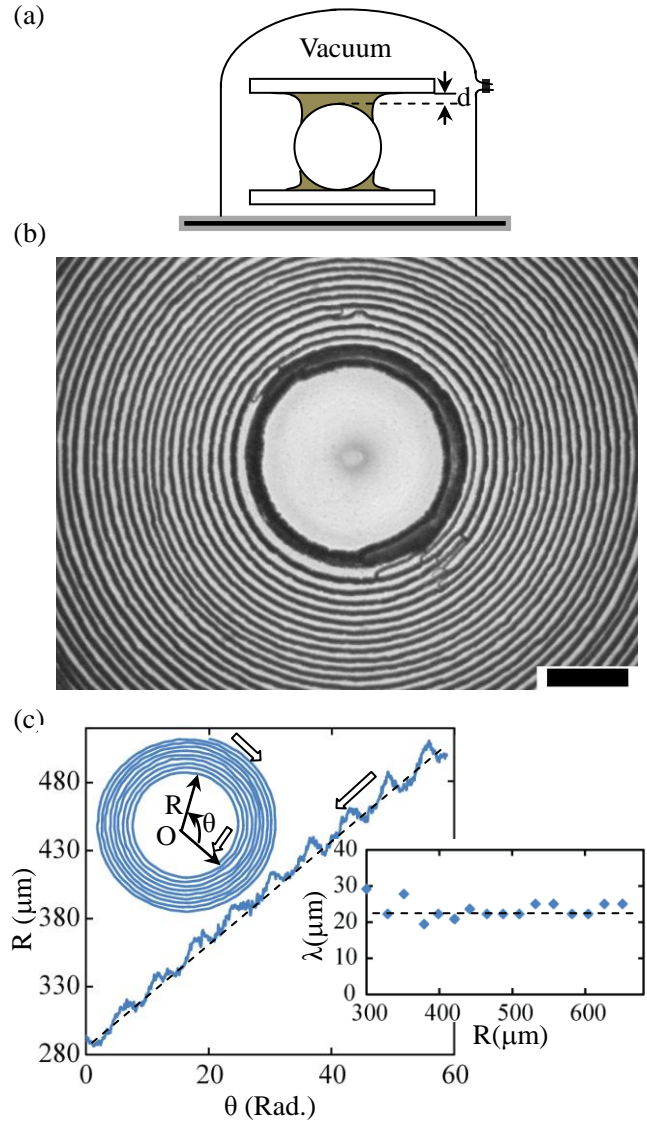

Figure 1 (Color online) The Archimedean spiral pattern with equi-spacing. (a) Schematic of the experimental setup. (b) A typical Archimedean spiral pattern. The scale bar is 200μm. (c) Plot of the radius R of the spiral stripe versus the angle θ in (b). R and θ are defined in the upper inset. The outlined arrow shows the direction of depinning propagation. The upper inset shows the track of the spiral in (b) and the lower inset shows the distance between neighboring stripes (wavelength λ) versus the radius. Dashed lines represent fitting to the data.

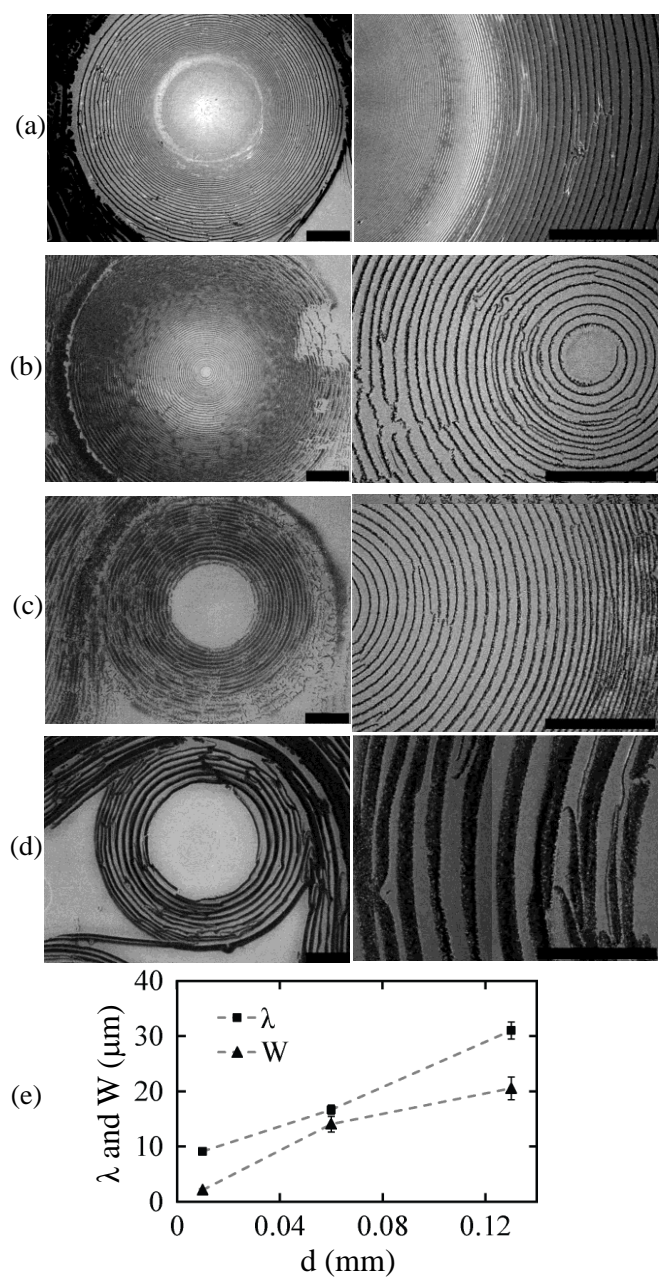

Figure 2 Dependence of the pattern on distance d. (a) d=0, (b) d=0.010mm, (c) d=0.060mm, (d) d=0.130mm. Figures on the right in (a), (b), (c) and (d) are magnified images of those on the right. (e) Plot of the wavelength (λ) and width of stripes (W) against distance d. The concentration of the solution used is 0.10 wt%. The scale bars are 200 μm (left) and 100μm (right).

$R = 289.87 + 3.75\theta$, which indicates an Archimedean spiral. The distance between stripes ($2\pi \times 3.75 = 23.55 \mu m$) agrees with the measurement in the inset of Fig. 1(c). The wavelength $\lambda$ is uniform, as shown in Fig. 1(c). The pattern deposited on the upper glass plate can be varied by changing the space between the top of the glass bead and the surface of the upper glass plate (represented by d in Fig. 1a). As shown in Fig. 2, the distance (wavelength $\lambda$) between neighboring stripes depends on the space d. When d=0, we obtain a gradient stripe pattern on the surface of the glass plate, as shown in Fig. 2a. When d≠0, there is an open space (without stripes) at the center of the pattern. When d≈0.010mm (Fig. 2b), the wavelength $\lambda$ is small ($\lambda = 9.14 \mu m$, average) and quasi-uniform. When d is larger, the wavelength $\lambda$ is large and uniform ($\lambda = 16.67 \mu m$, average, when d≈0.060mm and $\lambda = 31.02 \mu m$, average, when d≈0.130mm) (Fig. 2c &2d). A plot of the wavelength and the width of stripes is shown in Fig. 2e. Also, as d increases, the width of the stripes increases while the number of stripes decreases. The wavelength of the spiral pattern becomes uniform when d is adjusted to a suitable value. When d increases, the open space at the center of the pattern becomes large. These observations suggest that the morphology of this Archimedean spiral pattern can be controlled by tuning the distance d. The concentration of the solution has an important effect on the stripe pattern morphology, as shown in Fig. 3. With an increase in the concentration of the solution, both the wavelength and width increase while the number of stripes tends to decrease, as shown in Fig. 3b. Outside the spiral stripe pattern, a nonconcentric stripe pattern was found due to deposition via the evaporation of solution at the meniscus between the glass plates. The pattern formed a large area with a regular structure. We confirmed that a stripe pattern was found on the surface of the glass beads (not shown).

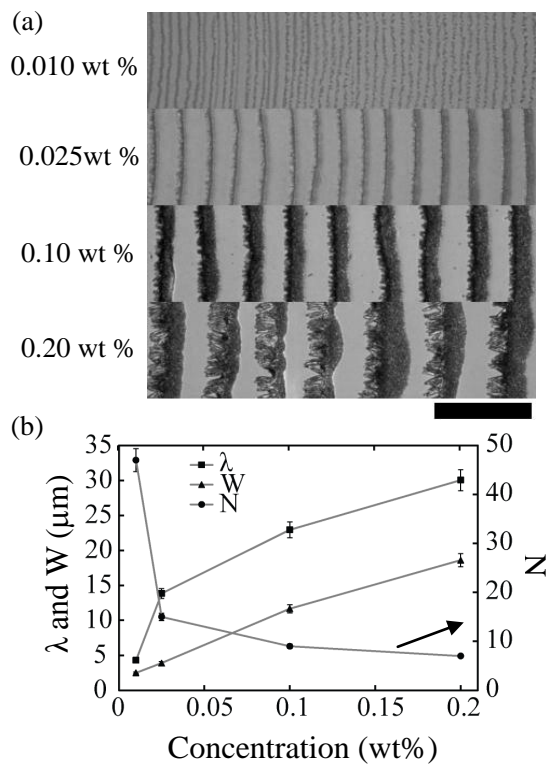

Figure 3 Effect of concentration on the stripe pattern. (a) Stripe patterns obtained with solutions of different concentrations. (b) Plot of wavelength (λ), width (W) and number (N) against the concentration of the solution in (a). The distance d between the bead and the glass plate is 0.040 mm. The scale bar is 50μm.

As shown in Fig. 1a, the solution was confined between two glass plates and surrounded the glass beads. During evaporation of the solution, Fullerene C60 molecules crystallized in the vicinity of the contact line by the "coffee ring" effect [19]. The deposition of Fullerene C60 crystal at the contact line pinned the contact line. The stripe pattern was formed by the stick-slip cycle of the contact line dynamics in the periphery of the solution surrounding the glass beads. When the contact line approached the edge of the glass beads, the meniscus was reshaped by capillary force. A new meniscus was formed between the glass beads and the glass plates. After evaporation of the solution at the new meniscus, an Archimedean spiral pattern is believed to be generated on the glass plates. An equi-spacing stripe pattern was found on the upper glass plate while a gradient pattern was found on the lower plate. By changing the distance d, we found that we could change the gradient pattern to an equi-spacing pattern. Magnified images of the stripe pattern with various values of distance d are shown in Fig. 2. When the glass bead was in contact with the glass plate (d=0), gradient stripes formed. The stripes become increasingly thinner as we move to the center of the pattern. We found many stripes with a gradient width and gradient wavelength, as shown in Fig. 2a. In contrast, as shown in Fig. 2b, Fig. 2c, and Fig. 2d, an equi-spacing stripe pattern was found with a desired number of stripes. In the region near the center of the pattern, an empty circular area remained. The wavelength and width of the stripes do not depend on the radius of the stripes, which reflects equi-spacing. The width of the stripes is about 10 μm, and more than 20 stripes were formed when the distance d≈0.060mm (Fig. 2c) (one stripe corresponds to one turn in the Archimedean spiral). About 10 stripes were formed when d≈0.130mm (Fig. 2d). The number of stripes (turns of the spiral) can thus be controlled by the distance d between the glass beads and the surface of the upper glass plate. When d is too large, the bridge between the glass plate and glass beads will break and no pattern is observed, as found in the experiment.

With regard to the mechanism of spiral stripe pattern formation, we consider the receding dynamics of the contact line due to evaporation of the solvent. Several factors, including capillary force (surface tension balance), the viscosity of the solution in the vicinity of the contact line, the pinning force from the substrate and the roughness of the surface, determine the motion of the contact line and the deposition of Fullerene C60 crystal on the substrate. Evaporation of the solvent at the air-solution interface in the vicinity of the contact line produces temperature and concentration gradients [19, 20]. The temperature gradient leads to a surface tension gradient along the air-solution interface and induces Marangoni flow. Evaporation in the vicinity of the contact line is much faster and causes a concentration gradient, which produces solvent flow toward the contact line, i.e., capillary flow [19, 20]. Marangoni flow and capillary flow transport Fullerene C60 molecules toward the vicinity of the contact line. The crystallization of solute pins the contact line. During pinning of the contact line, depinning of the contact line will occur when the depinning force (surface tensions at the air-solution interface and solution-substrate interface) counterbalances the maximum pinning force. Depinning deforms the surface of the meniscus and propagates along the contact line (Fig. 4a). As confirmed by snapshots taken with a high speed camera (Fig. 4b), the depinning front propagates along the pinning contact line through slipping motion of the contact line from the pinning position to the next pinning position. After one cycle of propagation, the depinning front begins a new cycle of propagation from the newly pinned contact line. The change in the radius with respect to the angle in Fig.1 (c) provides clear evidence of the propagation behavior of depinning. In the new cycle, the radius of the new stripe from depinning initially increases and then decreases because of evaporation of the solvent, which increases the slipping distance during depinning of the contact line. The propagation of the depinning front on the contact line is similar to a one-dimensional traveling wave in a diffusive system. The slipping motion of the contact line from the pinning position to the new position distorts the meniscus surface in the vicinity of the depinning front. The stress on the critically pinned contact line, from the distortion of the meniscus surface, causes the pinning of contact line to collapse. The distortion of the meniscus surface propagates in a manner similar to a distortional Rayleigh surface wave in an elastic medium and is governed by a Laplacian equation [21]. Thus, the depinning front propagates diffusively. The states of pinning and depinning correspond to the wet and dry states of the substrate where contact line undergoes slip motion. This phenomenon can be simply described as the transition between two stable states, which is described as the free energy profile with double minima. Thus, depinning propagation of the contact line can be described phenomenologically by the following equation:

$$\frac{\partial u}{\partial t} = D\frac{\partial^2 u}{\partial x^2} - \sigma u(u-a)(u-1) \qquad (1)$$

where $t$ is time, $x$ ($dx = rd\theta$) is the arc length along the contact line, $u$ is an order parameter that corresponds to a wet ($u=1$) or dry ($u=0$) substrate, $D$ is an apparent diffusion coefficient deduced from the free energy barrier of the slip motion in the original free energy functional [22], and $\sigma$ is a constant. Parameter $a$ ($a \in [0,1]$) represents the relative stability between $u=0$ and $u=1$ and depends on the degree of evaporation. The value of $a$ will increase from 0 toward 1 during the evaporation of the solvent. We set $\frac{1}{r^3}\frac{\partial r}{\partial \theta} \approx 0$ when $r$ is relatively large. Thus, equation (1) becomes

$$\frac{\partial u}{\partial t} = \frac{D}{r^2}\frac{\partial^2 u}{\partial \theta^2} - \sigma u(u-a)(u-1) \qquad (2)$$

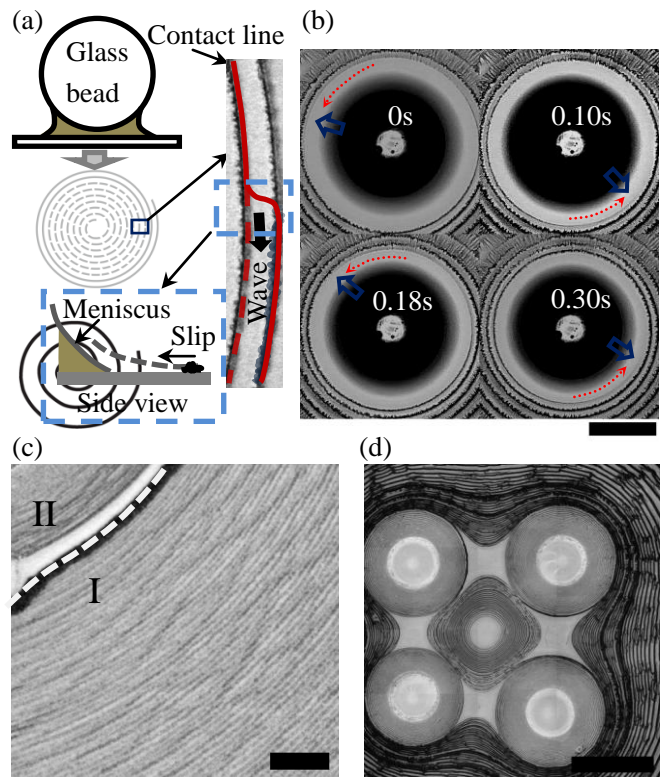

Figure 4 (Color online) Scenario of pattern formation. (a) Schematic of spiral formation via depinning propagation. (b) Snapshots of the traveling wave of the depinning front. Scale bar is 50 μm. The outlined and dotted arrows indicate the position and propagating direction of the depinning front at different times, respectively. (c) Stripe pattern from a solution confined between two parallel glass surfaces. Area I is the stripe pattern from the meniscus between two parallel glass surfaces and area II is the stripe pattern from the meniscus of a glass bead. The space between the two glass surfaces is 1.810mm and the concentration of solution is 0.025wt%. Scale bar: 100μm. (d) Self-organized pattern generated from the quasi-square arrangement of four beads. Scale bar: 1mm.

For a traveling wave, $u(\theta,t) = u(\theta - \varpi t)$ ($\varpi$ is the angular speed of the traveling wave). We can obtain $\varpi = \frac{\sqrt{D}}{r}$ from Eq. (2) [23]. In Eq. (2), the effect of evaporation (parameter $a$) dominates the difference in wavelength between two neighboring turns. According to our previous numerical simulation [6], the wavelength is dominated by the shape dynamics of the meniscus surface. The same profile dynamics of the meniscus surface will result in an identical wavelength. Let us now consider the meniscus that is formed between two approximately parallel surfaces when distance d is sufficiently large. The meniscus should have the same profile in the equilibrium state and the same critical depinning condition [6]. Thus, the critical condition (represented by a function of $a$) of evaporation for slip motion of the contact line is related to the evaporation time $t_0$ for one turn (assume a constant evaporation rate), the volume $V$ of the liquid bridge and the area $S$ of the meniscus surface. According to our previous analysis [6], we obtain

$$a_{critical} \sim \frac{t_0 S}{V} \approx \frac{2\pi \times 2\pi r d}{\varpi \pi r^2 d} \sim \frac{4\pi}{\sqrt{D}} \qquad (3)$$

The time evolution of wave propagation in each turn is governed by the same slip motion with the same critical condition from evaporation. A spiral pattern forms when depinning propagates on one side while the contact line on the other side is pinned. Under the same effect from evaporation, a uniform slip distance leads to the formation of an Archimedean spiral pattern, as found in the experiment.

The above model can be tested by an experiment in which the meniscus is confined between two parallel surfaces. Figure 4c shows the deposited stripe pattern of a Fullerene C60 crystal from the meniscus formed between two parallel glass surfaces. The periodic structure is continuous in the direction of motion of the contact line. As discussed previously [6], the meniscus dynamics determines the pattern morphology. The meniscus formed between a pair of parallel surfaces should have identical dynamic behavior during each stick-slip cycle. Thus, an equi-spacing stripe pattern should be found from repetitive stick-slip processes when a liquid column (the meniscus) is formed between parallel surfaces (Fig. 4c). In Fig. 1, the meniscus is formed between a flat surface and the surface of a glass bead. When distance d is sufficiently large, the shape of the meniscus between the two surfaces is similar to that between two parallel surfaces. Thus, an equi-spacing Archimedean spiral stripe pattern was formed when d was relatively large (Fig. 2c &Fig. 2d). The pattern transforms from a gradient pattern to an equi-spacing pattern when the distance d between the two surfaces increases, as shown in Fig. 2. The pattern can be controlled to show various morphologies by the arrangement of the glass beads (Fig. 4d). The arrangement of stripes can be conveniently controlled according to the intended function of the desired structure, such as in the design of solar cells [24], the development of microelectronic devices and data storage technologies [8].

In summary, we have demonstrated the self-organization of an Archimedean spiral pattern with a uniform wavelength. The wavelength and number of stripes can be controlled by tuning the space between the glass beads and the glass surface and by changing the concentration of the solution. This simple experiment should stimulate practical applications in device design, especially with regard to the fabrication of a micro electronic-coil by using a conductive material such as carbon nanotubes. We are extending our study along these lines.

This work was supported by the National Natural Science Foundation of China (No. 11147118) and the Japanese Ministry of Education, Culture, Sports, Science and Technology (MEXT) via Grants-in-Aid for Scientific Research (No. 20111002 & No. 23240044).

[1] C. Parneix, P. Vandoolaeghe, V. S. Nikolayev, D. Quéré, J. Li, B. Cabane, Phys. Rev. Lett. **105**, 266103 (2010).

[2] H. Bodiguel, F. Doumenc, B. Guerrier, Langmuir **26**, 10758 (2010).

[3] Z. L. Wang, R. R. Bao, X. J. Zhang, X. M. Ou, Ch. –S. Lee, J. C. Chang, X. H. Zhang, Angew. Chem. Int. Ed. **50**, 2811 (2011).

[4] D. Orejon, K. Sefiane, M. E. R. Shanahan, Langmuir **27**, 12834 (2011).

[5] J. Xu, J. F. Xia, Z. Q. Lin, Angew. Chem. Int. Ed. **46**, 1860 (2007).

[6] Y. –J. Chen, K. Suzuki, H. Mahara, T. Yamaguchi, Chem. Phys. Lett. **529**, 74 (2012).

[7] T. Kuykendall, P. J. Pauzauskie, Y. F. Zhang, J. Goldberger, D. Sirbuly, J. Denlinger, P. D. Yang, Nature Mater. **3**, 524 (2004).

[8] J. Lian, L. M. Wang, X. Ch. Sun, Q. K. Yu, R. C. Ewing, Nano Lett. **6**, 1047 (2006).

[9] P. J. Pauzauskie, D. J. Sirbuly, P. D. Yang, Phys. Rev. Lett. **96**, 143903 (2006).

[10] N. J. Suematsu, Y. Ogawa, Y. Yamamoto, T. Yamaguchi, J. Colloid Interface Sci. **310**, 648 (2007).

[11] E. Adachi, A. S. Dimitrov, K. Nagayama, Langmuir **11**, 1057 (1995).

[12] O. Karthaus, L. Grasjo, N. Maruyama, M. Shimomura, Chaos **9**, 308 (1999).

[13] H. Uchiyama, W. Namba, H. Kozuka, Langmuir **26**, 11479 (2010).

[14] H. Yabu, M. Shimomura, Adv. Funct. Mater. **15**, 575 (2005).

[15] A. Kuroda, T. Ishihara, H. Takeshige, K. Asakura, J. Chem. Phys. B **112**, 1163 (2008).

[16] K. Suzuki, T. Yamaguchi, Mol. Cryst. Liq. Cryst. **539**, 83 (2011).

[17] S. Watanabe, K. Inukai, S. Mizuta, M. T. Miyahara, Langmuir **25**, 7287 (2009).

[18] L. Frastia, A. J. Archer, U. Thiele, Phys. Rev. Lett. 106, 077801 (2011).

[19] R. D. Deegan, O. Bakajin, T. F. Dupont, G. Huber, S. R. Nagel, T. A. Witten, Nature **389**, 827 (1997).

[20] T Kajiya, E. Nishitani, T. Yamaue, M. Doi, Phys. Rev. **73**, 011601 (2006).

[21] H. Kolsky, *Stress Waves in Solids* (Dover Publications, New York, 1963).

[22] J. W. Cahn, J. E. Hilliard, J. Chem. Phys. **28**, 258 (1958).

[23] S. Fedotov, Phys. Rev. E **59**, 5040 (1999).

[24] S. E. Habas, H. A. S. Platt, M. F. A. M. van Hest, D. S. Ginley, Chem. Rev. **110**, 6571 (2010).